\documentclass[usenatbib,useAMS]{mn2e}
\usepackage{amssymb}
\usepackage{natbib}

\pdfoutput=1
\ifx\pdfoutput\undefined
 \usepackage{graphicx}
 \else
 \usepackage[pdftex]{graphicx}
 \fi

\def\beq{\begin{equation}}
\def\eeq{\end{equation}}

\title{Modelling the bow shock Pulsar Wind Nebulae propagating through a non-uniform ISM}

\author[O.D. Toropina et al.]
{O.D. Toropina,$^1$\thanks{E-mail:toropina@iki.rssi.ru}, M.M.
Romanova,$^2$, and R.V.E. Lovelace,$^{2}$\\
$^1$ Space Research Institute, Russian Academy of Sciences,
Profsoyuznaya 84/32, Moscow 117997, Russia\\
$^2$ Department of Astronomy, Cornell University, Ithaca, NY
14853-6801}

\date{\today}

\pagerange{\pageref{firstpage}--\pageref{lastpage}}
\pubyear{2017}

\begin{document}

\label{firstpage}

\maketitle

\begin{abstract}

Many pulsars propagate through the interstellar medium (ISM) with
supersonic velocities, and their pulsar winds interact with the
interstellar medium (ISM), forming bow shocks and magnetotails
(PWN). We model the propagation of pulsars through the
inhomogeneous ISM using non-relativistic axisymmetric
magneto-hydrodynamic (MHD) simulations. We take into account the
wind from the star, and the azimuthal and poloidal components of
the magnetic field, and investigate the PWN at different levels of
magnetization (the ratio of magnetic to matter energy-densities)
in the wind.
 We consider the interaction of  PWN with small-scale and large-scale imhomogeneities in the ISM
 at different values of magnetization.
 We conclude that the inhomogeneities in the ISM can change the shapes of the bow shocks and magnetotails
at different values of the magnetization.
 We compare the results of our simulations with the images of the
Guitar Nebula and other PWN that show irregularities in the shapes
of their bow shocks and magnetotails. We conclude that these
irregularities may be caused by the interaction of PWN with the
inhomogeneities in the ISM.

\end{abstract}



\section{Introduction}

Pulsars emit winds of relativistic particles and magnetic fields,
and are often surrounded by the Pulsar Wind  Nebulae (PWN) (e.g.,
\citealt{ReesGunn1974,KennelCoroniti1984}). Many pulsars have high
velocities and propagate supersonically through the ISM, and their
PWN interact with the ISM, forming bow shocks and magnetotails.
The bow shocks are often observed in the $H_\alpha$ spectral line
(e.g., \citealt{BrownsbergerRomani2014}). Many interesting
structures (bow shocks, very long tails, and jet-like features)
are observed in the X-ray (e.g., \citealt{KargaltsevPavlov2010})
and radio (e.g., \citealt{NgEtAl2010}) wavebands.

Pulsars have a wide range of velocities, $10~\rm{km
s}^{-1}\lesssim v \lesssim 1500~\rm{km s}^{-1}$, with two peaks in
their distribution at $v\approx 90~\rm{km s}^{-1}$ and $v\approx
500~\rm{km s}^{-1}$, so that many of them propagate supersonically
through the ISM \citep{ArzoumanianEtAl2002}.

\begin{figure*}
\begin{center}
\includegraphics[height=4.3cm]{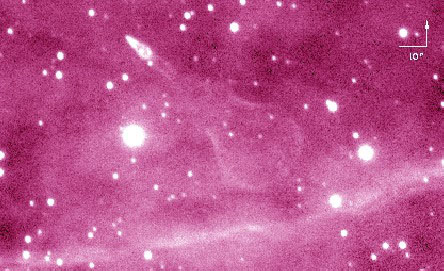}
\includegraphics[height=4.3cm]{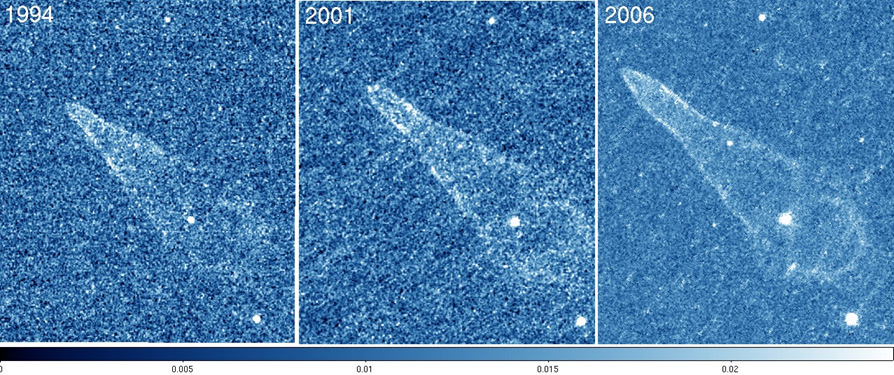}
\end{center}
\caption{\textit{Left panel:} The Guitar Nebula in $H_{\alpha}$,
imaged with the 5-meter Hale Telescope at the Palomar Observatory,
1995 \citep{CordesEtAl1993}. \textit{Right panel:} The head of the
Guitar Nebula in $H_{\alpha}$, imaged with the Hubble Space
Telescope in 1994, 2001, and 2006 \citep{GautamEtAl2013}. The
change in shape traces out the changing density of the ISM.}
\label{fig:guitar-obs}
\end{figure*}

One remarkable PWN is the Guitar Nebula, which is powered by the
pulsar PSR B2224+65 that travels at
a high velocity of about 1600 km/sec (see left panel of Figure
\ref{fig:guitar-obs}, from \citealt{CordesEtAl1993}). The
high-resolution observations in the $H_\alpha$ line, performed
with the Hubble Space Telescope in the years of 1994, 2001, and
2006, show that the shape of the Nebula's head
becomes wider with time (see right panels of Figure
\ref{fig:guitar-obs}, from \citealt{GautamEtAl2013}). The
variation in shape may be connected with the variation in the
density of the ISM
(\citealt{ChatterjeeCordes2002,ChatterjeeCordes2004}).
\citet{DolchEtAl2016} observed the Guitar Nebula with the 4-meter
Discovery Channel Telescope (DCT) at the Lowell Observatory in
2014. They compared the DCT observations with the 1995
observations by the Palomar 200-inch Hale telescope and found
changes in both the spatial structure and the surface brightness
of the nebula.

\begin{figure*}
\centering
\includegraphics[width=14cm,clip]{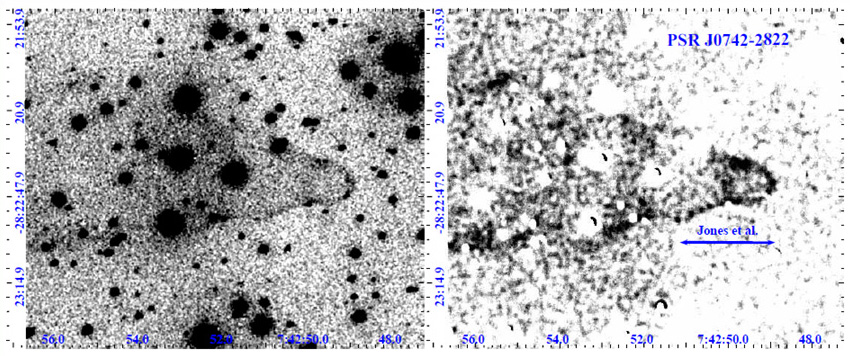}
\caption{PWN of the pulsar PSR J0742-2822, observed in the
$H_\alpha$ line \citep{BrownsbergerRomani2014}. \textit{Left
panel:} A median-filtered 3 x 600 W012 SOI image of PSR
J0742.2822, smoothed with a 0.45" Gaussian.
\textit{Right panel:} Same image, but with the scaled continuum
image subtracted and a 0.9" top-hat smoothing. The arrow indicates
the extent of the previous nebula detection.}
\label{fig:PSR-J0742}
\end{figure*}

Another interesting example of PWN is the $H_{\alpha}$ pulsar bow
shock connected with the pulsar PSR J0742-2822. Figure
\ref{fig:PSR-J0742} shows an image of PSR J0742-2822 taken by
\citet{BrownsbergerRomani2014}. This PWN
shows multiple irregularities in its shape, which suggests that
the pulsar may be travelling through small-scale fluctuations in
the ISM.

\begin{figure*}
\centering
\includegraphics[width=12cm,clip]{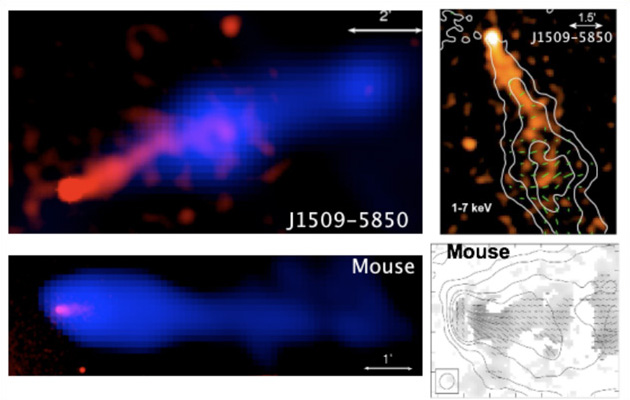}
\caption{X-ray and radio images of the very long pulsar tails, by
\citet{KargaltsevEtAl2012}. Right panels show the radio contours
and the direction of the magnetic field. The red and blue colours
in the left panels correspond to X-ray and radio, respectively.}
\label{fig:mouse-obs}
\end{figure*}

In the X-ray band, many of the PWN show irregularities in their
shapes, as well as very long tails (see review by
\citealt{KargaltsevEtAl2017}). Even longer tails are observed  in
the radio band (e.g., \citealt{NgEtAl2010}). Figure
\ref{fig:mouse-obs} shows two examples of PWN, observed in the
X-ray and radio bands: the PWN associated with the pulsar PSR
J1509-5850 (top panels), and the Mouse Nebula (bottom panels),
powered by the pulsar PSR J1747-2958 \citep{KargaltsevEtAl2012}.
One can see that both PWN have long tails in the X-ray band (red
colour), and even longer tails in the radio band (blue colour).
Polarization obtained in the radio band shows that the magnetic
field is mainly transversal in the case of PSR J1509-5850 (top
right panel), and mainly longitudinal in the Mouse PWN (bottom
right panel).
Both the heads of the PWN and their tails have different
irregularities in their shapes.

According to theoretical studies, a pulsar loses its rotational
energy in the form of the magnetic (Poynting flux) wind
\citep{GoldreichJulian1969,AronsTavani1994,Arons2004}. The
magnetic energy of this wind should be somehow converted to the
energy of the relativistic particles that interact with the ISM,
radiate  and form observed PWN (see, e.g.,
\citealt{ReesGunn1974,KennelCoroniti1984}). Different mechanisms
were proposed for conversion of the magnetic energy to the energy
of particles (e.g.,
\citealt{SironiSpitkovsky2011,LyubarskyKirk2001,Amato2014}). The
magnetization $\sigma$ (the relative amount of magnetic energy
flux compared to the energy flux of relativistic particles) varies
from $\sigma>>1$ near the light cylinder of the pulsar, to much
lower values at the shock front, where the pulsar wind interacts
with the ISM or with a supernova remnant. The early attempts to
built a theoretical model of the Crab nebula using the ideal
relativistic MHD approximation resulted in a conclusion that the
pulsar wind has to have $\sigma\sim 10^{-3}$ near its termination
shock \citep{ReesGunn1974, KennelCoroniti1984,BegelmanLi1992}. A
higher magnetization, $\sigma\sim 10^{-2}$, was later derived in
axisymmetric numerical simulations (e.g.,
\citealt{KomissarovLyubarsky2003,DelZannaEtAl2004,BogovalovEtAl2005}).
\citet{Komissarov2013} argued that the magnetization in the Crab
nebula can be much higher (than $\sigma\sim 10^{-2}$, as high as
$\sigma\sim 1$. Global 3D MHD simulations confirmed that the
magnetization may be high, $\sigma\gtrsim 1$
\citep{PorthEtAl2013,PorthEtAl2014}.

If a pulsar propagates supersonically through the ISM, then the
 PWN interacts with the ISM, forming a bow shock and magnetotails.
In the bow shock, the energy of accelerated particles  may
dominate over magnetic energy-density. However, in the
magnetotails, the magnetic energy-density may be  comparable to or
larger than the energy-density of the particles.   Long,
magnetically-dominated       magnetotails are expected to form in
the PWN \citep{RomanovaEtAl2005}. They may be visible, if the
accelerated particles propagate into the magnetotails, or
invisible otherwise.
 In modelling the supersonic PWN,
 it is important to
take into account both, the matter and the magnetic field
components of the PWN.

Supersonic propagation of pulsars through the ISM has been studied
in a number of axisymmetric non-relativistic and relativistic
hydrodynamic simulations (e.g.,
\citealt{Bucciantini2002,vanderSwaluwEtAl2003,GaenslerEtAl2004}).
Simulations have shown that the interaction of pulsar winds with
the ISM leads to the formation of several shocks, which have
similar properties in non-relativistic and relativistic
simulations. The bow shock can be approximately described by the
formulae derived theoretically by \citet{Wilkin1996}, though the
set of shocks and the dynamics of matter flow are more complex.

\citet{BucciantiniEtAl2005} performed axisymmetric relativistic
\textit{magnetohydrodynamic}  simulations of the supersonic
propagation of PWN through the ISM.   They have shown that a
similar set of shocks forms as in the hydrodynamic simulations of
the same problem. They studied the PWN at different values of the
magnetization parameter $\sigma$ in the wind. They took into
account the wind of relativistic particles and the azimuthal
component of the magnetic field.


Observations of the bow shock PWN show a variety of shapes, which
may be connected with
the interaction of the PWN with the inhomogeneities in the ISM
(e.g., \citealt{ChatterjeeCordes2004,BrownsbergerRomani2014}).
\citet{Wilkin2000} derived the shape of asymmetric bow shocks
interacting with the inhomogeneous medium analytically (where the
ISM has a density gradient).  \citet{VigeliusEtAl2007} studied
intrinsically-asymmetric PWN and the bow shocks interacting with
inhomogeneous medium numerically, using hydrodynamic 3D
simulations. They have shown that the shape of a bow shock changes
when a pulsar propagates through a region with a density gradient.
However, there is a variety of shapes of the PWN, which can also
be explained by the inhomogeneities in the ISM (e.g., small-scale
wiggles observed in different bow shocks in Figs.
\ref{fig:guitar-obs} - \ref{fig:mouse-obs}). On the other hand,
the long tails observed in some PWN in the X-ray and radio bands
may be connected with the propagation of relativistic particles
along the long magnetotails (e.g., \citealt{RomanovaEtAl2005}).

The goals of our paper are twofold. On one hand, we  study the
propagation of the PWN through the ISM taking into account both
toroidal and poloidal components of the magnetic field, and the
properties of the bow shocks and magnetotails at different
magnetizations in the pulsar wind. On the other hand, we
investigate the interaction of the PWN with the inhomogeneities in
the ISM.

\begin{figure*}
\begin{center}
\includegraphics[width=12cm,clip]{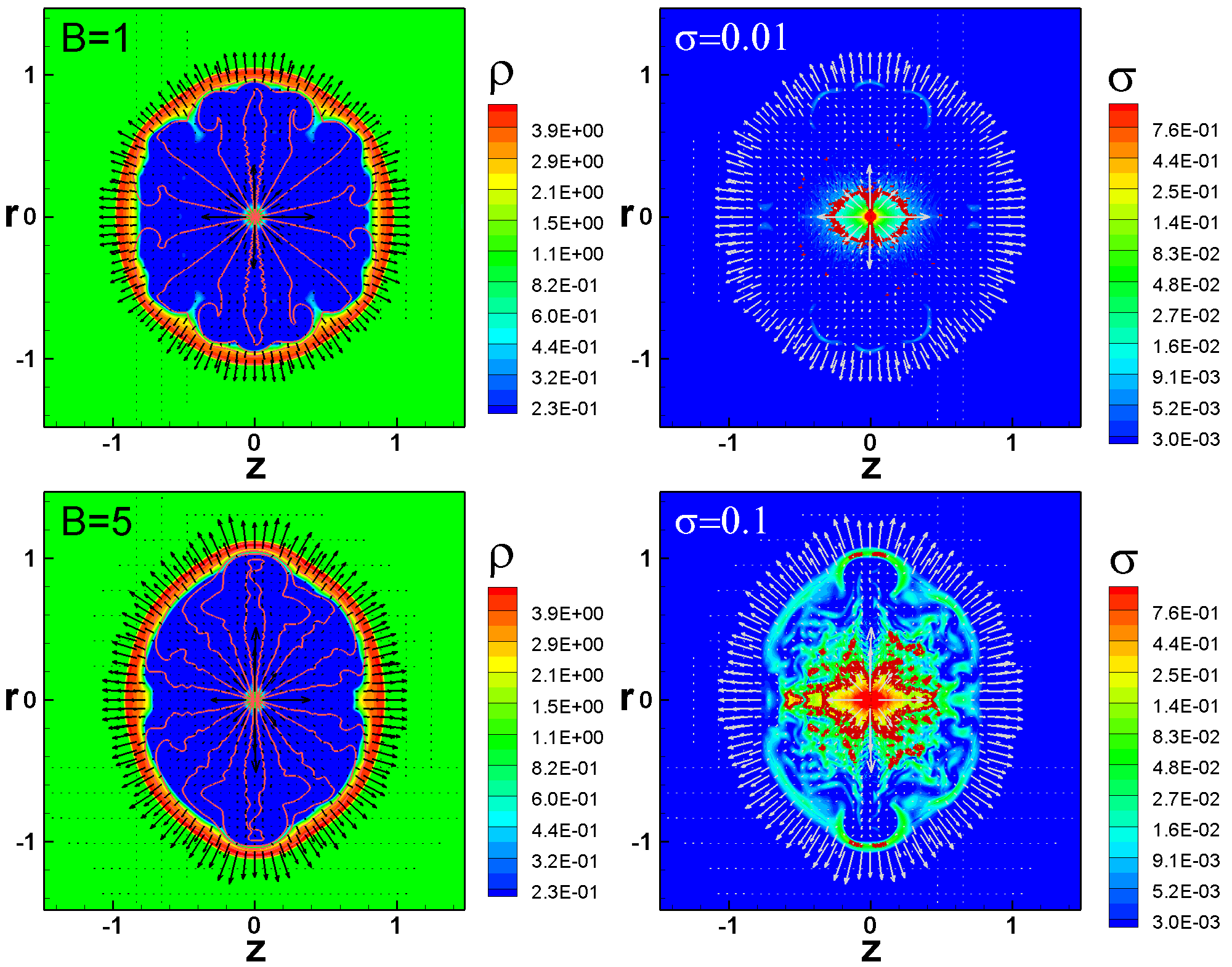}
\end{center}
\caption{\textit{Left panels:} Distribution of density (colour
background) and poloidal magnetic field (black lines) in the wind
around a non-moving star in the cases of weaker, $B=1$ (top
panel), and stronger, $B=5$ (bottom panel), magnetic fields. The
arrows are proportional to matter flux. The density varies between
$\rho_w=1$ near the star and 0.01 in the middle of the bubble.
\textit{Right panels:} Distribution of $\sigma$. The red line
shows values of $\sigma=0.01$ in the top plot, and $\sigma=0.1$ in
the bottom plot. } \label{fig:pwn-rho-sigma-4}
\end{figure*}

\begin{table}
\centering
\begin{tabular}{l|llllll}
 Model        & Description & $M$  & $M_w$ & $B$ &  $\Omega $  &    $\sigma$   \\
\hline
$B1M20w50$    & low $\sigma$         &  20   & 50   & 1 &    1  &   0.01    \\
\hline
$B5M20w50$     & medium $\sigma$   &  20   & 50   & 5 &    1  &    0.1    \\
\hline
$B5M20w0$      & high $\sigma$   &  20    & 0    & 5 &    0  &    $> 1$     \\
\hline
$B1M50w50$     & Guitar Nebula   &  50   & 50   & 5 &    1  &    0.1    \\
\hline
\end{tabular}
\caption{Parameters of the main simulation models.  Here, $M$ is
the Mach number of the pulsar, $M_w$ is the Mach number of the
matter in the wind at the base of the flow, $B$ and $\Omega$ are
the magnetic field and angular velocity of the star in
dimensionless units, and $\sigma$ is the typical maximum value of
magnetization in the magnetotail.} \label{tab:models}
\end{table}

\section{Numerical Model}

We performed MHD simulations to investigate the supersonic
propagation of a wind-ejecting magnetized neutron star through the
uniform and  non-uniform ISM. We used an axisymmetric,
non-relativistic resistive MHD code. The code incorporates the
methods of local iterations \citep{ZhukovEtAl2012} and
flux-corrected transport \citep{BorisBook1973}. The flow is
described by the resistive MHD equations
\citep{LandauLifshitz1960}:

$$
    {\partial \rho \over
    \partial t}+
    {\bf \nabla}{\bf \cdot}
\left(\rho~{\bf v}\right) =0{~ ,}
$$
$$
\rho\frac{\partial{\bf v}} {\partial t}
              +\rho ({\bf v}
\cdot{\bf \nabla}){\bf v}=
    -{\bf \nabla}p+
{1 \over c}{\bf J}\times {\bf B} + {\bf F}^{g}{ ~,}
$$
$$
    \frac{\partial {\bf B}}
{\partial t} =
    {\bf \nabla}{\bf \times}
\left({\bf v}{\bf \times} {\bf B}\right)
    +
    \frac{c^2}{4\pi\sigma_e}
\nabla^2{\bf B} {~,}
$$
$$
    \frac{\partial (\rho\varepsilon)
}{\partial t}+
    {\bf \nabla}\cdot \left(\rho
\varepsilon{\bf v}\right) =
    -p\nabla{\bf \cdot}
{\bf v} +\frac{{\bf J}^2} {\sigma_e}{~.} \eqno(1)
$$

We assume axisymmetry $(\partial/\partial \phi =0)$, but calculate
all three components of velocity ${\bf v}$ and magnetic field
${\bf B}$.
    We consider the equation of state for an ideal gas,
$p=(\gamma-1)\rho \varepsilon$, where $\gamma=5/3$ is the specific
heat ratio and $\varepsilon$ is the specific internal energy of
the gas. The gravitational force ${\bf F}^{g} = -GM\!\rho{\bf
R}/\!R^3$.
      The equations incorporate Ohm's law, ${\bf
J}=\sigma_e({\bf E}+{\bf v}  \times {\bf B}/c)$, where $\sigma_e$
is the electric conductivity.
     The associated magnetic diffusivity,
$\eta_{\rm m} \equiv c^2\!/(4\pi\sigma_e)$, is assumed to be
constant. Diffusivity is important at the sites of reconnection of
the magnetic field lines. In this study, we take a small
diffusivity. It is only slightly larger than the numerical
diffusivity, which is low at our high grid resolution (see
\citealt{ToropinaEtAl2001} for an analysis of diffusivity in our
models).

We use a cylindrical, inertial coordinate system
$\left(r,\phi,z\right)$, with the $z-$axis parallel to the star's
dipole moment ${\mu}$ and rotation axis ${\bf \Omega}$. The vector
potential $\bf A$ is calculated so that the condition ${\bf
\nabla}\cdot{\bf B}=0$ is satisfied at all times. We rotate the
star at an angular velocity ${\Omega}$. The intrinsic magnetic
field of the star is taken to be an aligned dipole, with vector
potential ${\bf A}={\mu} \times{\bf R}/{R^3}$. A detailed
description of the numerical model can be found in
\citet{ToropinaEtAl2001,ToropinaEtAl2006,ToropinaEtAl2012}.

We measure length in units of the Bondi radius \citep{Bondi1952},
$R_B \equiv {GM}/c_{s}^2$, where $c_s$ is the sound speed at
infinity. The size of the computational region is $R_{\rm
max}=1.2$, $Z_{\rm min}=-0.8$ and $Z_{\rm max}=3.2$ in units of
the Bondi radius.  The radius of the numerical star (inner
boundary) is $R_s=0.025$. The grid $N_R\times N_Z$ is $385\times
1281$ in most cases.

 We measure velocity $v$ in units of the pulsar's
velocity, $v_p$, and time in units of  $t_0=(Z_{max} -
Z_{min})/v_p$, which is the crossing time of the computational
region with the pulsar's velocity, $v_p$.  The density is measured
in units of density of the interstellar medium, $\rho_0$, and the
magnetic field strength is measured in units of $B_0$,
 which is
the field
determined at the distance $r=0.25 R_s$. The conversion from
dimensionless to dimensional variables is described in
\citealt{ToropinaEtAl2001}.

\begin{figure*}
\begin{center}
\includegraphics[width=18cm,clip]{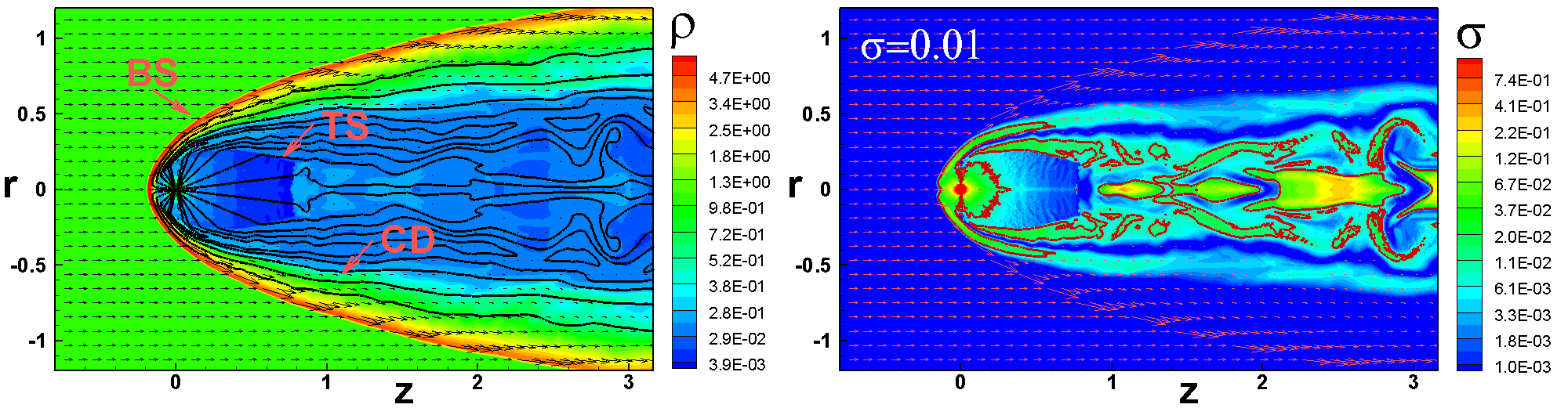}
\end{center}
\caption{\textit{Left panel:} An example of a shock wave that
forms around a PWN with low $\sigma$ (model $B1M20w50$). The
background represents the logarithm of density. The solid lines
are magnetic field lines. The arrows are proportional to matter
flux. The main shocks are shown with red arrows, where TS stands
for Termination Shock, BS - for Bow Shock, and CD - for Contact
Discontinuity. \textit{Right panel:} Same, but the background
shows the logarithm of $\sigma$, and the red line shows the value
of $\sigma=0.01$.} \label{fig:b1m20-shock-sigma}
\end{figure*}

\begin{figure*}
\begin{center}
\includegraphics[width=18.0cm,clip]{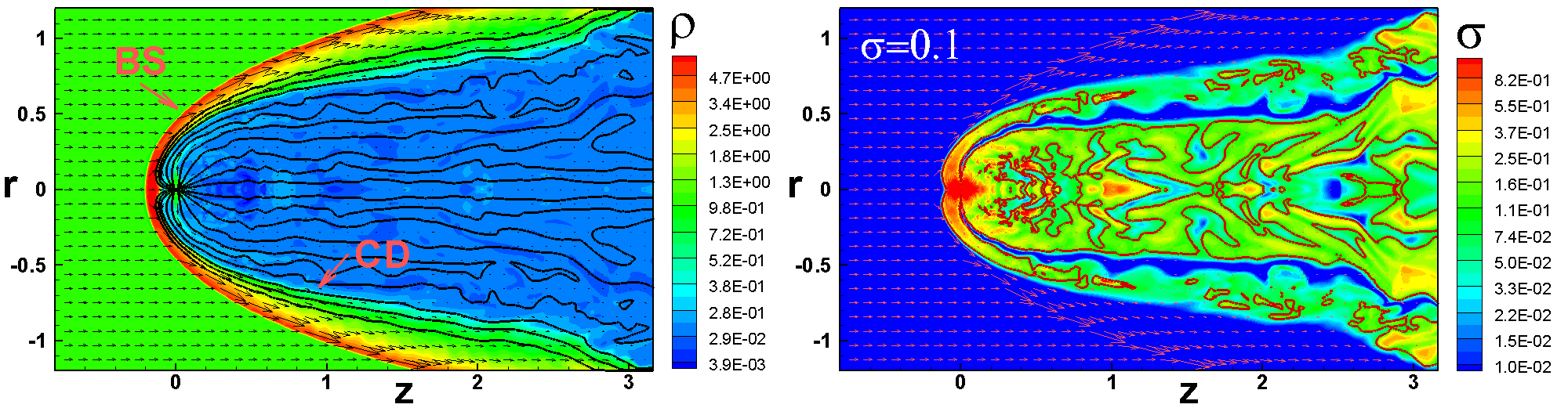}
\end{center}
\caption{\textit{Left panel:} An example of shock wave which forms
around PWN with medium $\sigma$ (model $B5M20w50$). The background
represents the logarithm of density. The solid lines are magnetic
field lines. Arrows are proportional to matter flux. \textit{Right
panel:} Same, but the background shows the logarithm of $\sigma$,
and the red line shows the value of $\sigma=0.1$.}
\label{fig:b5m20-shock-sigma}
\end{figure*}

\begin{figure*}
\begin{center}
\includegraphics[height=5.cm,clip]{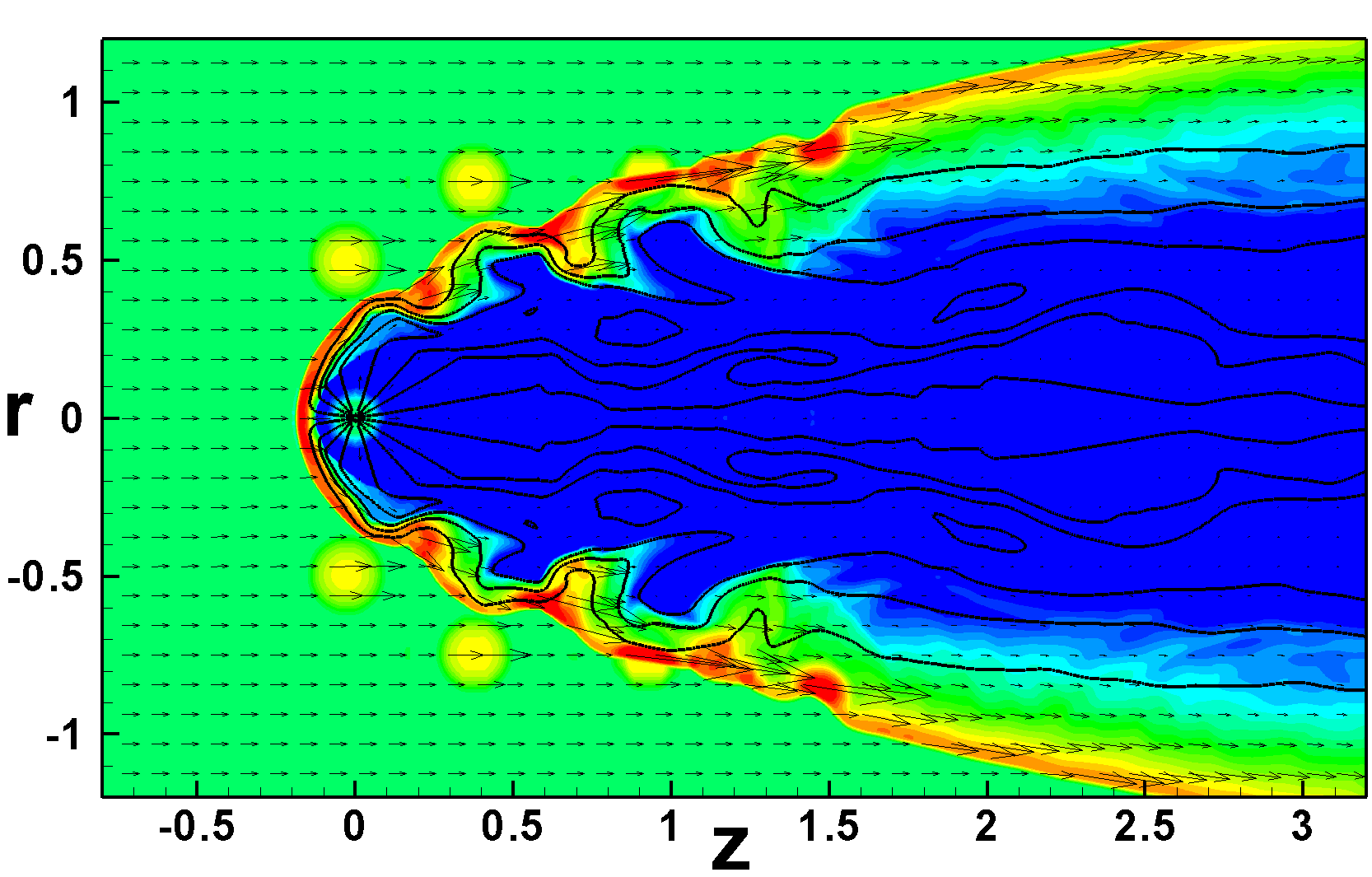}
\includegraphics[height=5.cm,clip]{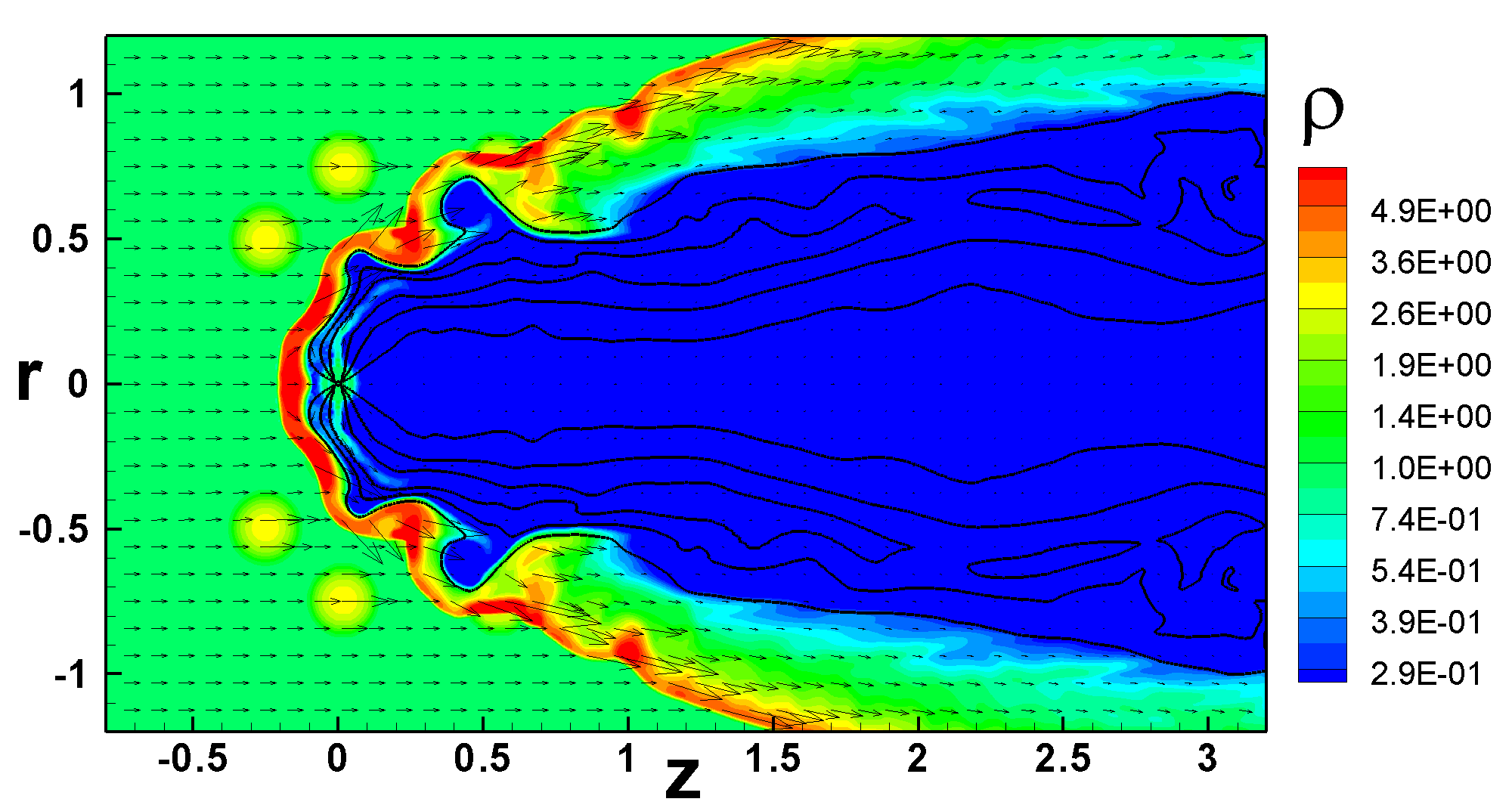}
\end{center}
\caption{Interaction of the bow shock with  small-scale clouds of
maximum density $\rho_{\rm cloud}=3\rho_0$  in the model
$B1M20w50$ with low $\sigma$ (left panel) and in the model
$B5M20w50$ with medium $\sigma$ (right panel). The background
represents the logarithm of density. The solid lines are magnetic
field lines.} \label{fig:chess-2}
\end{figure*}

\section{Results of simulations}

\subsection{Modelling PWN around non-moving pulsars}

As a first step, we studied the PWN in the case where the pulsar
has  zero velocity. A pulsar's winds represent the relativistic
flow of particles and a predominantly azimuthal magnetic field
(e.g., \citealt{ReesGunn1974,KennelCoroniti1984}). The wind
originates at the light cylinder and propagates to much larger
distances, where it interacts with a supernova remnant or with the
ISM (e.g., \citealt{KaspiEtAl2006}). The magnetization varies from
large values near the light cylinder to smaller values at larger
distances from the star. In our study, we are interested in the
external parts of the wind, which interact with the ISM. This is
why we suggest that a neutron star and its light cylinder are
located deeply inside our inner boundary (numerical star), and we
model only the external regions of the PWN. As soon as many
parameters of the PWN are not known, such as the magnetization, we
suggest that the wind should incorporate some matter and magnetic
field. We suggest that the poloidal component of the magnetic
field should also be present, and consider a general case where
matter and both components of the field are present in the wind.

To model the pulsar wind, we
placed a ring about our numerical star at radii $1.5 R_s < R_w <
2.2 R_s$ and generated a radial wind from this ring. The wind has
density $\rho_w$ and poloidal velocity $v_w$, corresponding to the
Mach number $M_w=v_w/c_s=50$ (where $c_s$ is the sound speed in
the ISM). We also rotate the ring with an angular velocity of
$\Omega$. The star has a dipole magnetic field. The axis of the
dipole is aligned with the symmetry axis. The kinetic energy of
matter {flowing from the ring}  is higher than the magnetic
energy, so that the magnetic field is stretched out by the blowing
wind. It is also twisted by the azimuthal component of the wind.
The ratio between the poloidal and the azimuthal components is
regulated by the value of $\Omega$. This way, we obtain the flow
of matter and the magnetic field, including both azimuthal and
poloidal components.

Our wind is non-relativistic. Experiments with different
velocities and densities in the wind have shown that the modelled
PWN has similar characteristics between the cases of very
low-density,  high-velocity winds (at $M_w=200-1,000$), and
higher-density, lower-velocity winds with the same kinetic energy.
We chose the latter approach because, in magnetohydrodynamics,
simulations are much longer in the presence of low densities. We
chose $M_w=50$ and $\rho_w=1$ (equal to the reference density in
the ISM) in our reference models.

We varied the magnetic field at the surface of the numerical star
so as to obtain different levels of  magnetization in the wind.
 We measure the magnetization in the flow using the
non-relativistic version of $\sigma$:
\begin{equation}
\sigma = \frac{B^2/8\pi}{p+\rho (v_r^2 + v_z^2 + v_\phi^2)} .
\end{equation}
Here, $p$ is the local gas pressure. We investigate PWN which have
different levels of the magnetization. For our representative
runs, we took two values of the magnetic field at the surface of
the numerical star: $B=1$ and $B=5$, which provided lower and
higher levels of magnetization, respectively.

Fig. \ref{fig:pwn-rho-sigma-4} shows examples of PWN in the cases
of $B=1$ (top panels) and $B=5$ (bottom panels). The nebula
expands into an ISM of constant density. One can see that the
shock wave in case of lower field is more spherical, compared with
the case of a higher field. The density decreases rapidly, as
$\rho\sim (r/R_w)^{-2}$,  from $\rho=\rho_0$ at $r=R_w=0.055$ up
to $6\times 10^{-4}$ at $r=1$.  At $B=1$, the shock wave  slightly
differs from the spherical one due to the action of the magnetic
field. At $B=5$, the shock wave becomes more elongated in the
equatorial  direction. The right panels show the magnetization.
Red line corresponds to $\sigma=0.01$ in case of $B=1$ (top panel)
and $\sigma=0.1$ in case of $B=5$ (bottom panel). Therefore, we
call the  model with $B=1$ as the `low magnetization' model, and
the model with $B=5$, the ``high magnetization" model.  Nebulae
continue to expand into the ISM, so that we show only an
intermediate time during expansion, with the goal to demonstrate
our numerical PWN for a non-moving star.

\begin{figure*}
\begin{center}
\includegraphics[width=14cm,clip]{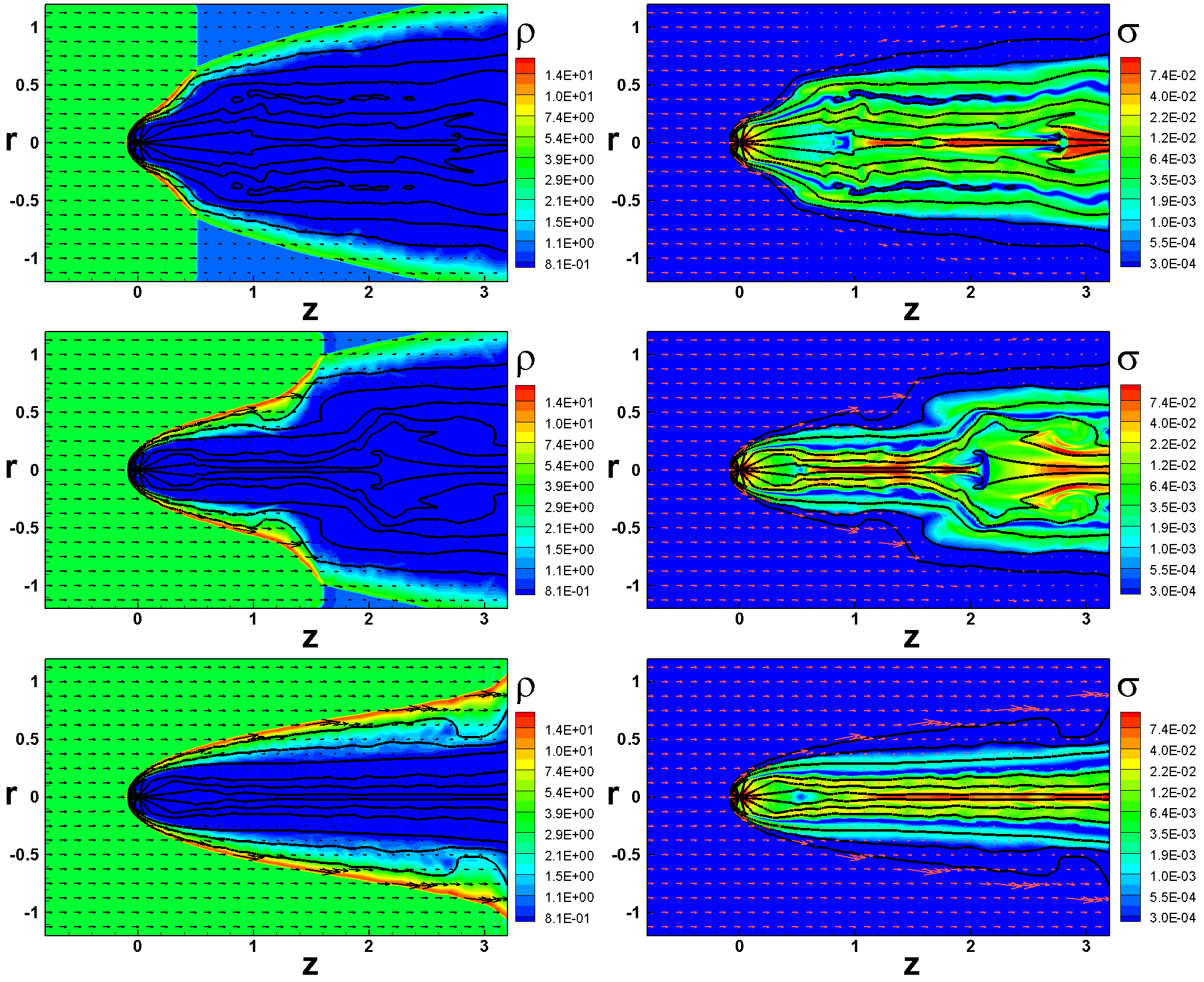}
\end{center}
\caption{\textit{Left panels:} The propagation of a PWN of low
magnetization (model $B1M20w50$) through a large cloud of density
of $\rho_{\rm cloud}=3 \rho_0$ at different moments in time. The
background represents the logarithm of density. The solid lines
are magnetic field lines. \textit{Right panels:} Same, but the
background shows magnetization $\sigma$.} \label{fig:b1-cloud-l-6}
\end{figure*}

\begin{figure*}
\begin{center}
\includegraphics[width=14cm,clip]{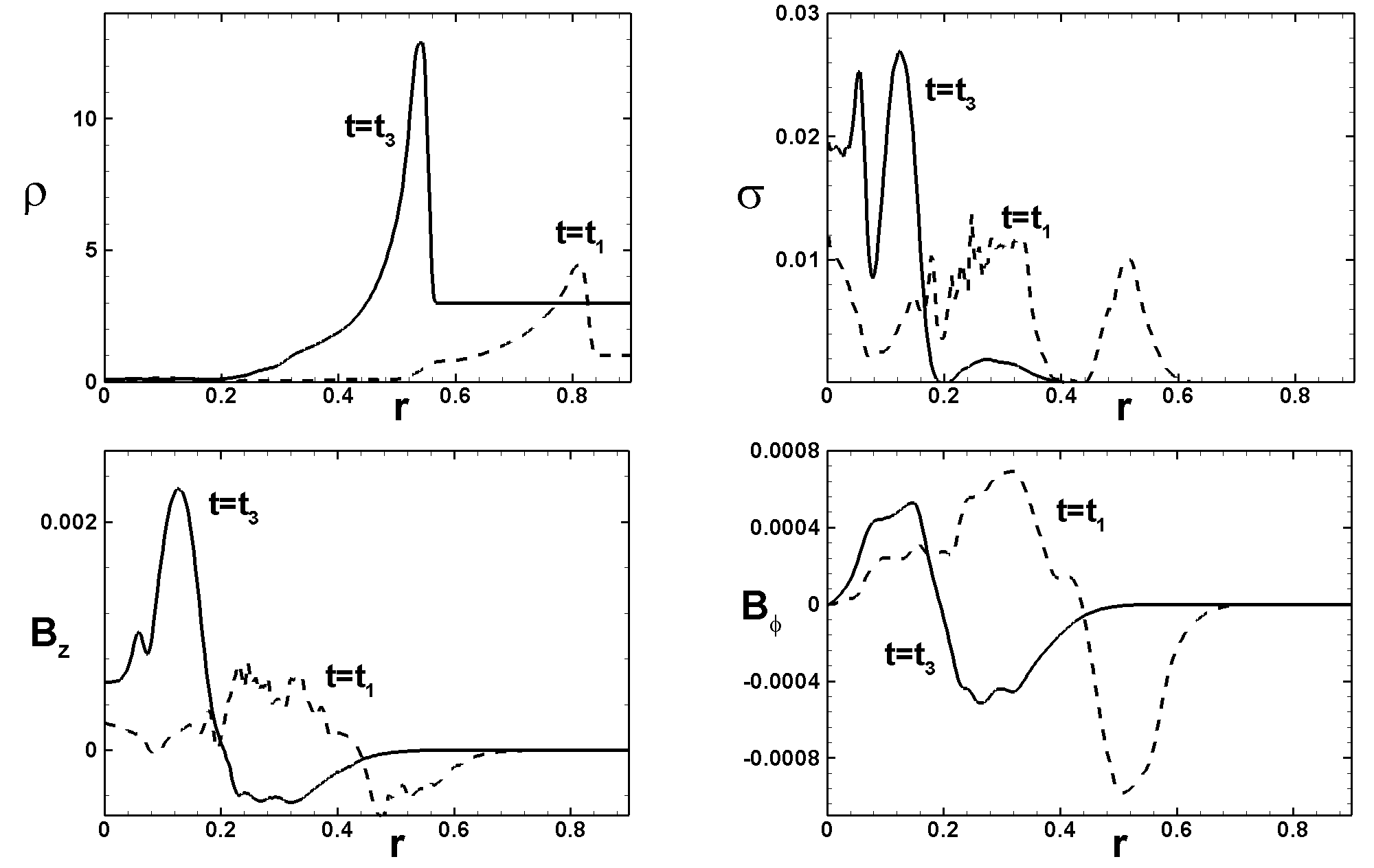}
\end{center}
\caption{Distribution of different parameter value across the bow
shock at $z=1$ in the case where a pulsar and its bow shock move
through a large cloud (see Fig. \ref{fig:b1-cloud-l-6}). Time
$t=t_1$ corresponds to the undisturbed bow shock (top panel of
Fig. \ref{fig:b1-cloud-l-6}), while time $t=t_3$ corresponds to
the new bow shock (bottom panel of the same figure).  Dashed and
solid lines correspond to moments $t=t_1$ and $t_3$, respectively.
Top panels show density and $\sigma$ distribution, while the
bottom panels show distribution of the $z-$ and $\phi-$ components
of the magnetic field. } \label{fig:1d-b1-cloud-4}
\end{figure*}

\subsection{Propagation of PWN through an ISM of constant density}

As a next step, we investigate the propagation of a magnetized PWN
through the ISM. Instead of moving the star through the simulation
region, we fixed the star and moved the ISM with Mach number $M$.
In our representative runs, we took Mach number $M=20$ and the
Mach number in the wind $M_w=50$ . We observed that the PWN formed
a set of shocks which were earlier observed in different
hydrodynamic and MHD simulations (e.g.,
\citealt{Bucciantini2002,vanderSwaluwEtAl2003}). Fig.
\ref{fig:b1m20-shock-sigma} shows an example of matter flow in the
model with relatively low magnetization, $\sigma\approx 0.01$
(model $B1M20w50$). One can see: (a) a bow shock (BS), where the
ISM matter is stopped by the PWN, (b) the bullet-shaped
termination shock (TS), where the wind from the star interacts
with the ISM matter, and the contact discontinuity (CD) shock,
where the matter that passed through the BS interacts with the
matter that passed through the TS. The stand-off distance of the
bow shock can be calculated from the balance of the ram pressure
in the stellar wind and the ISM: $\rho_w v_w^2=\rho_0 v_0^2$. The
density of the wind from the star (from the ring) decreases with
distance as $\rho_w=\rho_0 (r/R_w)^{-2}$, where $R_w=2.2 R_s =
0.055$ in our dimensionless units. Taking into account the fact
that the velocity of matter in the ISM is $v_0=M c_{s0}$ and the
velocity of matter at the base of the ring is $v_w=M_w c_{s0}$, we
obtain the stand-off distance:
$$
R_{\rm sd} =R_w \frac{M_w}{M} \approx 0.14 \frac{(M_w/50)}{(M/20)}
~.
$$
Fig. \ref{fig:b1m20-shock-sigma} shows that the bow shock has a
stand-off distance of $R_{\rm sd}\approx 0.14$.

Fig. \ref{fig:b5m20-shock-sigma} shows a similar plot for the case
of higher magnetization, $\sigma\approx 0.1$ (model $B5M20w50$).
One can see that there is no bullet-shaped PWN, but instead the
PWN wind propagates to larger distances along the poloidal field
lines of the magnetotail. We suggest that this is the result of
relatively high magnetization in the flow.  Earlier,
\citet{BucciantiniEtAl2005} concluded that in their case of high
magnetization, $\sigma=0.2$, the flow is governed by the magnetic
field, and the shape of the termination shock differs from the
shapes in cases of the lower magnetization. In our model, the
magnetization is also high, $\sigma\sim 0.1$, and also the
poloidal component is present, which influenced the structure of
matter flow in the magnetotail.

 The stand-off distance of the bow
shock, $R_{\rm sd}$, is approximately the same as in the case of
lower magnetization (in model $B1M20w50$), because,  although
 the magnetic field is stronger, but matter energy-density is still
larger than the magnetic energy-density.

\begin{figure*}
\begin{center}
\includegraphics[width=16cm,clip]{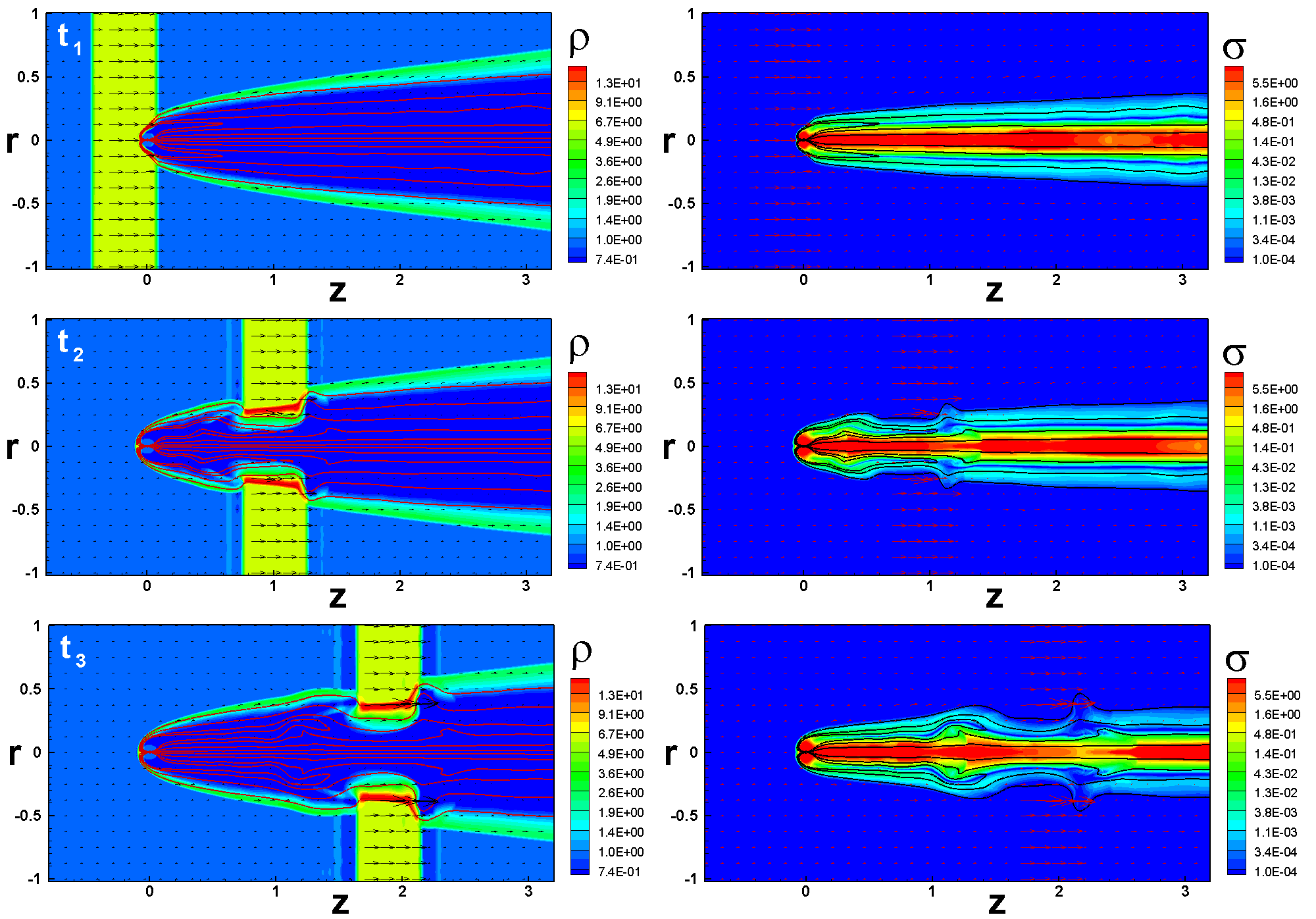}
\end{center}
\caption{Propagation of a cylindrically-shaped cloud  through the
strongly-magnetized PWN (model $B5M20w0$) at three moments in
time: $t_1, t_2$ and $t_3$. The Mach number ${\cal M}=20$ and the
density of the cloud $ {\rho} / {\rho}_0 = 6 $.
The background in the left panels represents the logarithm of
density. The solid lines are magnetic field lines. The length of
the arrows is proportional to the poloidal matter flux. The
background in the right panels represents the logarithm of
magnetization, $\sigma$.} \label{fig:b5w0-stripe-6}
\end{figure*}

\begin{figure*}
\begin{center}
\includegraphics[width=14cm,clip]{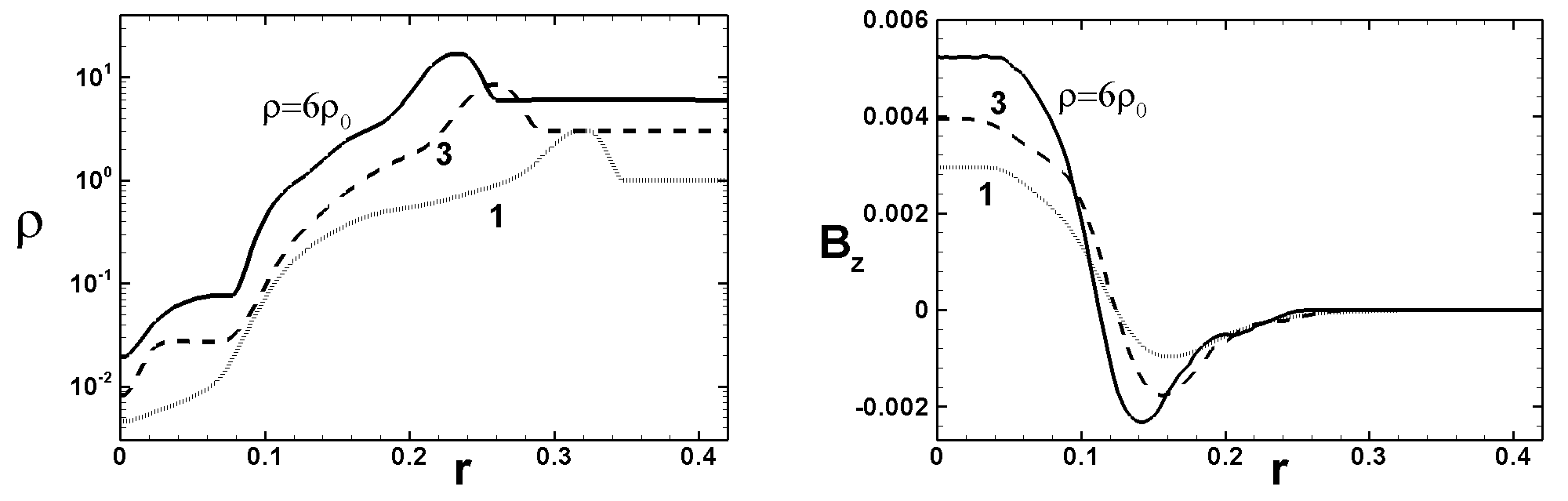}
\end{center}
\caption{\textit{Left panel}: The radial distribution of density
across the magnetotail  in the region where the cloud passes
through the magnetotail at a distance of $z=0.5$ from the star.
\textit{Right panel}: Same, but for the distribution of the
$B_z-$component of the magnetic field.  The solid, dashed and
dotted lines represent the cloud densities $ {\rho} / {\rho}_0 =
6, 3$ and 1, respectively.} \label{fig:1d-d5w0-2}
\end{figure*}

\subsection{Propagation of PWN through the inhomogeneous ISM}

Next, we investigate the propagation of a PWN through the ISM with
an inhomogeneous matter distribution. Observations point to large
and small-scale inhomogeneities, and therefore we consider two
types of imhomogeneities: (1) small-scale clouds, whose size is
comparable with the width of the bow shock
at the place of interaction, and (2) large-scale clouds, which are
much larger than the size of the bow shock.

We model small-scale inhomogeneities as a set of small clouds with
the Gaussian distribution of density, with the maximum density of
$\rho_{\rm max}$
and the half-width of the Gaussian at $\Delta r=0.2$. To keep the
pressure balance between the clouds and the rest of the ISM, we
take the temperature in the cloud to be $T_{\rm cloud}=T_{\rm
ISM}*\rho_0/\rho_{\rm cloud}$.

We experimented with clouds of different densities, and found that
the shape of the bow shock varies significantly if the density in
the cloud is $\rho_{\rm cloud} \gtrsim 2 \rho_0$. At $\rho_{\rm
cloud} = 2 \rho_0$, a wavy structure starts to become visible,
while at $\rho_{\rm cloud} = 3 \rho_0$, the bow shock changes its
local shape significantly.

Fig. \ref{fig:chess-2} shows a wavy variation of density in the
bow shocks that appears after the propagation of clouds of density
$\rho_{\rm cloud} = 3 \rho_0$ in the models with low and medium
magnetizations ($B1M20w50$ and $B5M20w50$, respectively). Such an
interaction with small-scale inhomogeneities may explain the
wiggles in the shape of the bow shock observed in  PSR J0742-2822
(see Fig. \ref{fig:PSR-J0742}).

Fig. \ref{fig:b1-cloud-l-6} shows an example of propagation of a
large cloud of density $\rho_{\rm cloud}=3 \rho_0$ through the PWN
in the model ($B1M20w50$) with a lower magnetization. The left
panels show that the cloud compresses the bow shock  and, in the
final state, when the star enters the cloud completely, the bow
shock has a smaller opening angle (the Mach cone). Our cloud is in
pressure balance with the rest of the ISM, so that the sound speed
in the cloud is three times lower than in the rest of the ISM.
Therefore,  the Mach number of the star inside the cloud is
 $M_{\rm new}= v_p /c_s=3
M_{\rm old}=60$. This is why we observed a smaller opening angle
of the Mach cone. It is interesting to note that the interaction
of the bow shock with the smaller-scale clouds can also be
interpreted this way. The right panels of Fig.
\ref{fig:b1-cloud-l-6} show that the magnetic flux becomes
compressed.

Fig. \ref{fig:1d-b1-cloud-4} compares different quantities across
the magnetotail at the distance of $z=1$ from the star at the
moment of time $t=t_1$, corresponding to the top panels of Fig.
\ref{fig:b1-cloud-l-6} (before entering the cloud), and at
$t=t_3$, corresponding to the bottom panels of Fig.
\ref{fig:b1-cloud-l-6} (after entering  the cloud). The top left
panel shows that, after entering the cloud, the density in the bow
shock is a few times larger. The magnetization also becomes larger
(see top right panel of the same figure). The $B_z$ component of
the magnetic field increases by a few times (bottom left panels),
while the azimuthal component, $B_\phi$ (which was originally on
the order of the $B_z-$ component), slightly decreases and becomes
a few times smaller than the $B_z$  component. These plots show
that the bow shock becomes more narrow, while the magnetotail
becomes even narrower with a stronger poloidal field. This example
demonstrates that the interaction with the ISM may lead to an
enhancement of the magnetic field, which may possibly be a reason
for the re-brightening of the magnetotail in X-ray
\citep{KargaltsevEtAl2017}.


\subsection{Propagation of strongly-magnetized PWN through inhomogeneous ISM}

In the above examples, the magnetotails are matter-dominated.
Here, we consider the case where a significant part of the
magnetotail is magnetically-dominated. To model this situation, we
take a  model similar to  $B5M20w50$, but suggest that there is no
pulsar wind, $v_w=0$ (model $B5m20w0)$). In this case, a
strongly-magnetized star propagates through the ISM, and the ISM
matter stretches its magnetosphere, forming a strongly-magnetized
magnetotail. The top panels of Fig. \ref{fig:b5w0-stripe-6} show
the distributions of density (left panel) and magnetization (right
panel) in the beginning of interaction with the cloud. The red
colour in the right panel shows the region where $\sigma>1$. One
can see that the magnetotail has a high magnetization and
stretches out to large distances from the star.

Next, we investigate the interaction of strongly-magnetized PWN
with the inhomogeneities in the ISM.  We take a cloud of density
$\rho_{\rm cloud}=6 \rho_0$, which has the shape of a cylinder of
width $\Delta z\approx 0.5$, and let this cloud move towards the
bow shock and interact with it. Top panels of Fig.
\ref{fig:b5w0-stripe-6} show the density distribution (left panel)
and $\sigma$ distribution (right panel) in undisturbed bow shock
at time $t=t_1$. Other panels show the density and $\sigma$
distribution at two moments in time ($t_2$ and $t_3$) when the
cloud passes through the bow shock. One can see that the
magnetotail is compressed and a limb-brightening appears (as in
the cases of weaker-magnetized magnetotails). The panels show that
the magnetic flux is compressed in the region of the cloud.

We also calculated a  model with a lower density in the cloud, $
{\rho} / {\rho}_0 = 3 $, and compared the density  distributions
with those of the higher-density cloud, $ {\rho} / {\rho}_0 = 6 $,
and with the case of a homogeneous ISM, $ {\rho} / {\rho}_0 = 1 $.
The left panel of Figure \ref{fig:1d-d5w0-2} shows the density
distribution across the magnetotail at a distance of $z=0.5$ and
at time $t=0.33 t_0 $ (the time is taken to be the same for all
three models). The dotted line represents a uniform medium with
constant density ${\rho}_0$.  The dashed line represents the cloud
density $ {\rho} / {\rho}_0 = 3$, and the solid line represents
the cloud density $ {\rho} / {\rho}_0 = 6 $. The density maximum
corresponds to $r=0.32, 026$ and $0.23$, for these three cases.
Therefore,  the tail of the magnetosphere becomes narrower, when
the density in the cloud increases.

The right panel of Figure \ref{fig:1d-d5w0-2} shows the
distribution of the $B_z$ component of the magnetic field across
the magnetotail at the same distance, $z=0.5$.
The dotted line represents a uniform medium with constant density
${\rho}_0$. The dashed line represents the cloud density $ {\rho}
/ {\rho}_0 = 3 $, and the solid line represents the cloud density
$ {\rho} / {\rho}_0 = 6$. One can see that the magnetic field is
largest in the case of the cloud of higher density.

\begin{figure*}
\begin{center}
\includegraphics[width=14cm,clip]{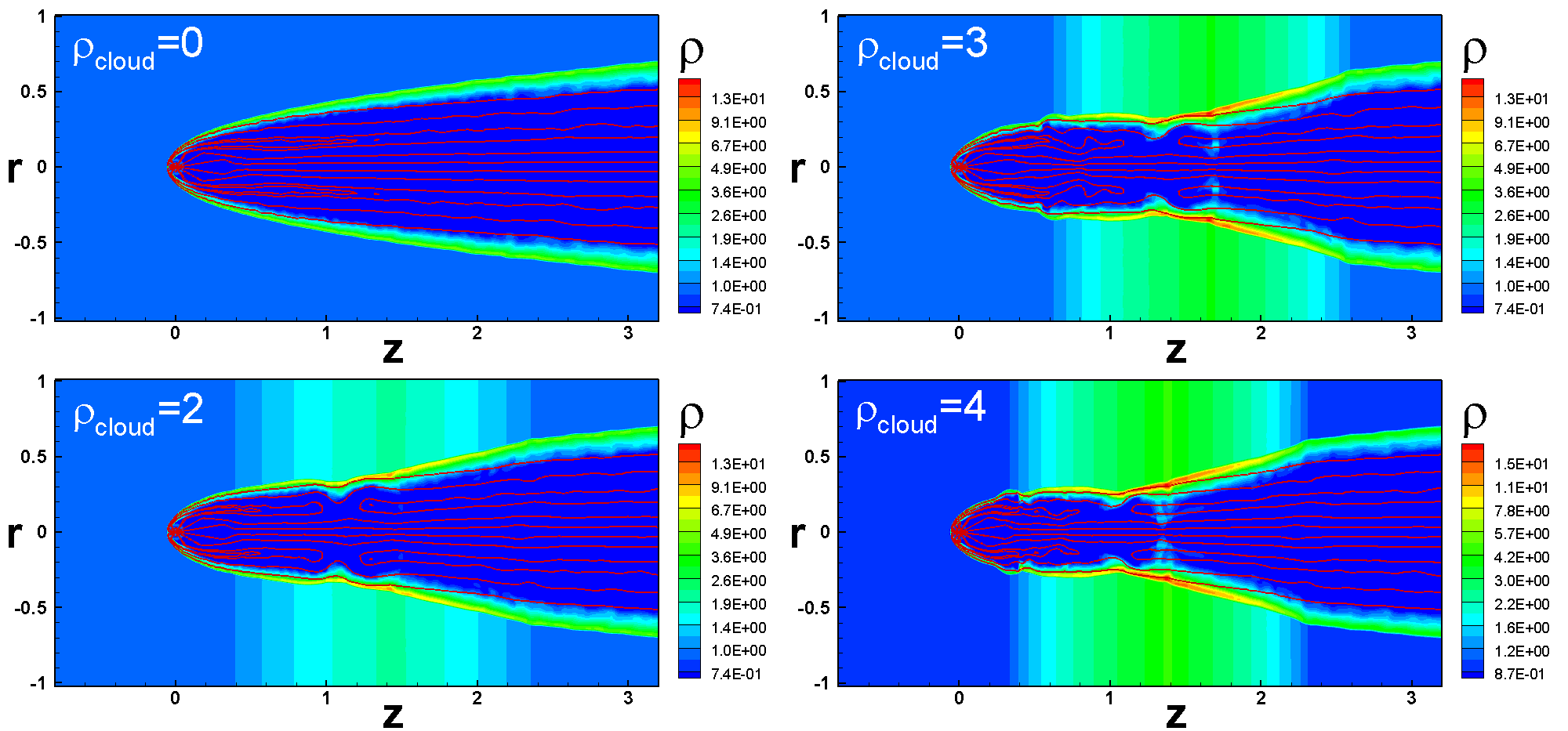}
\end{center}
\caption{Modelling
the head of the Guitar Nebula   bow shock at different density
gradients in the cloud. The Mach number of the star is $M=50$
(model $B1M50w50$). The density in the cloud has a positive
gradient in the z-direction up to the middle of the cloud, and
then the negative density gradient at larger distances. The
maximum density in the cloud increases with the density gradient
and corresponds to  $\rho_{\rm cloud}/\rho_0=1,2,3,4$ for $k_{\rm
grad}=0,1,2,3$, respectively .} \label{fig:guitar-4}
\end{figure*}

\subsection{Modelling the Guitar Nebula}

 Fig. \ref{fig:guitar-obs} shows that in the years of 2001 and 2006, the
head of the bow shock of the Guitar Nebula was observed to expand,
compared with the observations of year 1994
\citep{ChatterjeeCordes2002,ChatterjeeCordes2004,GautamEtAl2013}.
In addition, in the years of 2001 and 2006, the head of the nebula
was observed to have changed morphologically, showing new regions
where the opening angle of the bow shock is nearly zero. That is,
the opposite sides of the bow shock are almost parallel to each
other. This part of the bow shock also shows higher levels of
radiation in the $H_\alpha$ spectral line.
\citet{ChatterjeeCordes2004} suggested that the pulsar entered the
low-density region (so that the head expanded) with a positive
density gradient, so that parts of bow shock become compressed,
providing the lower opening angle of the Mach cone (see also
\citealt{VigeliusEtAl2007,MorlinoEtAl2015}).

We modelled the propagation of a bow shock through the density
gradient using cylindrical cloud
of width $z_1<z<z_2$. Experiments  have shown that, to match the
observations, the density in the cloud should initially increase,
and then decrease. We chose the following density distribution:
\begin{eqnarray} \rho_{\rm cloud}=\rho_0[1+k_{\rm
grad}(z_1-z)(z_c-z_1)] ~~~ at ~~ z_1<z<z_c ~, \nonumber\\
 \rho_{\rm cloud}=\rho_0[1+k_{\rm
grad}(z_2-z)(z_2-z_c)]~~~ at ~~ r_c<r<r_2, \nonumber
\end{eqnarray}
 where $z_c$ is
the position of the middle of the cloud, $z_c=(z_2-z_1)/2$, and
$k_{\rm grad}$ is the gradient.

Fig. \ref{fig:guitar-4} shows the results of our simulations. We
took a model with parameters similar to those of the model
$B1M20w50$, but with a higher Mach number, $M=50$, so as to model
the small Mach cone observed in the head of the Guitar Nebula. The
top left panel shows the bow shock before its interaction with the
cloud. The other panels show the bow shock after its  shape has
been modified by the cloud. The plots correspond to the density
gradients $k_{\rm grad}=1, 2$ and $3$. The maximum density in the
cloud is equal to to $\rho_{\rm cloud}=2, 3$ and $4$,
respectively. One can see that, in the right two panels, there is
a part of the bow shock with a Mach cone that is  approximately
equal to zero.  The density is enhanced in this
cylindrically-shaped region, as in the observations of the Guitar
Nebula's head in the years of 2001 and 2006. This density
enhancement may explain the brightening of those parts of the bow
shock in the Guitar Nebula. We should note that only this type of
cloud (with a smoothly decreasing density towards the edges)
provides a good match. For example, the passage through a much
larger cloud with a density gradient will not provide the observed
shape.

\citet{ChatterjeeCordes2004} also suggested that the widening of
the head of the Guitar Nebula observed in the year of 2001,
compared with the observations of year 1996, may be connected with
its entering the region of 0.7 times lower density. In 2006, the
head appeared to be even wider, so it may enter an ISM of even
lower density. We decreased the density in the ISM by the factor
of  $0.7$ and $0.5$, and observed that the head of the bow shock
becomes larger and also the Mach cone becomes wider, as expected.
We agreed that this may be a possible reason for the widening of
the cone in the Guitar Nebula.


\section{Conclusions}

We performed MHD simulations of the bow shock PWN propagating
through a uniform and non-uniform ISM
 at three levels of magnetization.
Our main findings are the following:

\textbf{1.} The interaction of the bow shock with the small-scale
inhomogeneities in the ISM leads to a wavy structure in the bow
shock. The amplitude of the ``waves" in the bow shock increases
with the ratio  $\rho_{\rm cloud}/\rho_0$. For example, variation
in the shape of the bow shock of pulsar PSR J0742-2822 can be
explained by the propagation through a series of clouds with a
density ratio of $\rho_{\rm cloud}/\rho_0\approx 3$.

\textbf{2.} The interaction of the bow shock with a large-scale,
dense cloud leads to the compression of the bow shock and the
formation of a new bow shock with a smaller opening angle. In the
opposite scenario, where the bow shock passes a cloud of  lower
density than that of the ISM, then the opening angle increases.

\textbf{3.} The shape of the head of the Guitar Nebula (the
density cylinder) can be explained by the density gradient in the
cloud, if the gradient coefficient is  $k_{\rm grad}\gtrsim 2$.
The passage of the cloud of lower density may explain the
expansion of the bow shock in the Guitar Nebula (as suggested
earlier by \citealt{ChatterjeeCordes2004}).

\textbf{4.} The compression of the magnetotail by the cloud leads
to a higher level of magnetization and the stronger magnetic field
in the magnetotail. The passage of the cloud through the
magnetotail may lead to a local amplification of the magnetic
field and to a re-brightening of the magnetotail in the X-ray at
large distances from the star (as suggested earlier by
\citealt{KargaltsevEtAl2017}).

\textbf{5.} In the cases of strongly-magnetized magnetotails, the
passage of the cloud also leads to: (a) the local compression of
the magnetotail, (b) a wavy structure, and (c) a local enhancement
of the magnetic field, which can lead to a re-brightening of the
tail in X-ray.

\section*{Acknowledgments}

We thank Dr. V.V. Savelyev for the development of the original
version of the MHD code and Alisa Blinova for help. This work was
supported by RFBR grant 17-02-00760 and by Russian Academy of
Science program I.28P. MMR and RVL were partially supported by
NASA through \textit{Chandra} Award  GO3-14082C. Resources
supporting this work were provided by the NASA High-End Computing
(HEC) Program through the NASA Advanced Supercomputing (NAS)
Division at the NASA Ames Research Center and the NASA Center

\end{document}